# Synthèse Comportementale Sous Contraintes de Communication et de Placement Mémoire pour les composants du TDSI

Gwenolé Corre, Philippe Coussy, Pierre Bomel, Eric Senn, Eric Martin

Email : prénom.nom@univ-ubs.fr
LESTER LAB, UBS University, CNRS FRE 2734

**Résumé** – La conception de systèmes complexes en traitement de l'image et du signal implique de réduire les coûts architecturaux et de maximiser les performances temporelles tout en considérant les contraintes de communication et d'accès mémoire durant la conception et l'intégration d'accélérateurs matériels dédiés. Malheureusement, les blocs matériels utilisés dans les flots de conception semi-automatiques traditionnels n'autorisent pas une flexibilité suffisante pour garantir cet objectif. Dans cet article, nous présentons une méthodologie et un outil qui autorisent la synthèse d'applications en traitement du signal et de l'image sous contraintes de communication et de mémorisation. Basé sur un ensemble de modèles formels, notre outil GAUT aide le concepteur à trouver un compromis entre performance et complexité architecturale.

***Abstract*** – *The design of complex Digital Signal Processing systems implies to minimize architectural cost and to maximize timing performances while taking into account communication and memory accesses constraints for the integration of dedicated hardware accelerator. Unfortunately, the traditional Matlab/ Simulink design flows gather not very flexible hardware blocs. In this paper, we present a methodology and a tool that permit the High-Level Synthesis of DSP applications, under both I/O timing and memory constraints. Based on formal models and a generic architecture, our tool GAUT helps the designer in finding a reasonable trade-off between the circuit's performance and its architectural complexity. The efficiency of our approach is demonstrated on the case study of a FFT algorithm.*

## 1. Introduction

En raison de la complexité des applications de traitement du signal numérique (TDSI), les concepteurs ont besoin d'un chemin plus direct entre la spécification et l'implémentation. Des flots de conception et les outils de CAO associés sont nécessaires pour maîtriser la complexité des systèmes et réduire les temps de conception. Ceci a mené au développement d'environnements d'aide à la conception explorant l'espace de solutions. Dans [1], [2] et [3] les auteurs proposent des approches qui utilisent des outils tels Matlab/Simulink/Stateflow [4] pour la spécification de l'application et qui produisent automatiquement une architecture VHDL RTL du système. Basés sur des macro-générateurs qui utilisent les mécanismes "generic"/"generate", la synthèse matérielle de l'application peut être résumée à une instanciation de blocs. Toutefois, bien que ce type de composant puisse être paramétrable, il repose sur un modèle architectural fixé dont la personnalisation reste limitée. Ce manque de flexibilité est particulièrement vrai pour (1) l'interface de communication dont les ordres d'acquisition et de production des données d'entrée/sortie ainsi que leurs caractéristiques temporelles (débit) sont définies et (2) pour l'unité de mémorisation dans laquelle la localisation des données est figée. La synthèse de haut niveau peut être utilisée pour résoudre ce manque de flexibilité. Quelques outils comme SystemC Compiler et Monet introduisent des modèles pour les E/S et la mémoire. Ils restent toutefois limités par la complexité des algorithmes à synthétiser [5], [6] et par la non dissociation de la spécification des contraintes de communication et de la spécification du traitement algorithmique.

Dans le domaine du temps réel et des applications utilisant de grande quantité de données, les ressources de calcul et de mémorisation doivent intégrer l'augmentation du flux de données. La conception d'architecture ou de système doit se concentrer sur (1) la réduction des goulets d'étranglement sur les bus et les tampons d'entrées/sorties, (2) la réduction du coût de stockage des données et (3) le respect de contraintes temporelles strictes en terme de cadence et de latence. La conception doit s'appuyer sur une modélisation des contraintes d'E/S et de mémoires de données internes, une analyse des contraintes pour vérifier l'existante d'une solution architecturale et l'utilisation de la synthèse de haut niveau pour générer l'architecture du composant.

Dans [7] et [8], nous proposons une méthodologie de conception de SoC basée sur la ré-utilisation d'IP algorithmique. Nous utilisons la synthèse de haut niveau sous contrainte d'E/S pour fournir des IP prenant en compte les contraintes d'intégration d'un système complet (cadence d'arrivée des données, technologie, format des bus …). Dans [9], nous introduisons une nouvelle approche prenant en compte l'architecture des mémoires interne et le placement des données lors de la synthèse de haut niveau. ***Dans cet article, nous proposons un flot de conception basé sur des modèles formels qui autorise la synthèse de haut niveau d'applications TDSI sous contrainte de communication et de mémorisation. Le concepteur peut ainsi spécifier les E/S, la distribution et le placement des données en mémoire interne et la cadence comme contraintes pour la synthèse de composants matériels. Cette approche peut être intégrée au flot de conception Matlab/Simulink pour augmenter la flexibilité des blocs matériels.***

Dans la section 2, nous formulons le problème de la synthèse de haut niveau sous contraintes d'E/S et de mémorisation. La section 3 présente les principales étapes de notre approche et les modèles formels qui y sont associés. Dans la section 4, nous démontrons l'intérêt de notre approche avec un exemple de synthèse d'une transformée de Fourier Rapide (FFT).

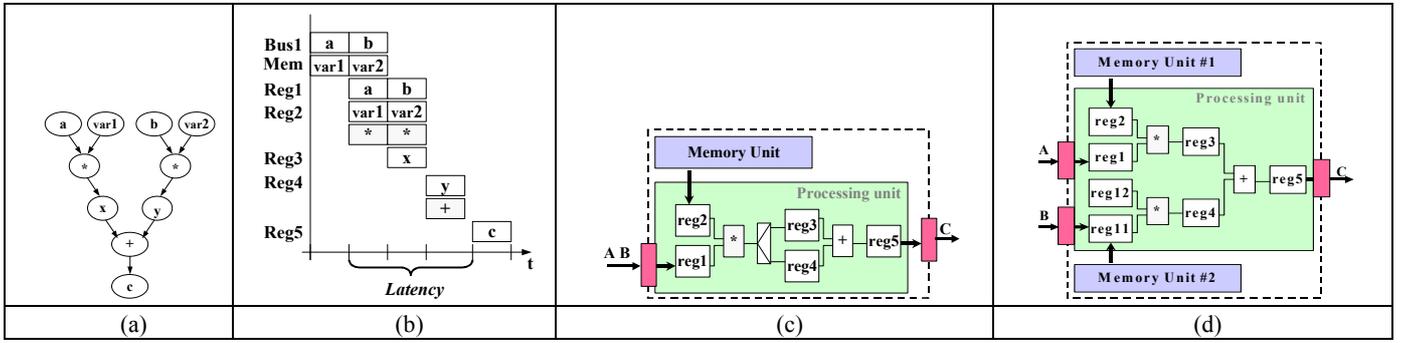

Fig. 1: (a) Signal Flow Graph SFG, (b) Comportement temporelle, (c) Architecture séquentielle, (d) Architecture parallèle

## 2. Formulation du problème

Considérons un composant matériel basé sur une architecture générique composée de deux unités fonctionnelles principales, une unité de traitement *UT* et une unité mémoire *UM*. Soit l'expression $c = (a*var1)+(b*var2)$ où *var1* et *var2* sont des variables stockées en mémoire. La figure 1 (a) montre le Graphe flot de Signaux SFG de cette expression. Ce composant reçoit les données d'entrée *a* et *b* provenant via le *bus1* et retourne le résultat *c* sur le *bus2*. Toutes les E/S consommées et produites par l'unité de traitement sont lues et produites suivant un ordre de séquence fixe $S = (a,b,c)$, càd $t_a < t_b < t_c$. La séquence de lecture des variables internes est complètement déterministe, càd $t_{var1} < t_{var2}$, avec $t_{var1} = t_a$ et $t_{var2} = t_b$. Dans ce contexte, un seul banc mémoire est suffisant pour satisfaire les contraintes temporelles décrites dans la figure 1 (b) où la latence est de 2 cycles. La figure 1 (c) présente une architecture de composant qui intègre 1 multiplieur, 1 additionneur et 5 registres.

Considérons maintenant la séquence de transfert de données suivante $S_{busses} = (a | b, c)$ càd $t_a = t_b < t_c$. Si la latence requise pour produire le résultat est suffisamment grande ($\geq 3$ cycles), elle permet de réordonner (sérialiser) les données d'entrée *a* et *b*. La solution architecturale incluant un banc mémoire peut être réutilisée. Cependant, cette solution nécessite l'ajout d'un adaptateur (wrapper) d'entrée composé d'un registre, d'un multiplexeur et d'un contrôleur. Si la latence est faible ($\leq 2$ cycles), le concepteur doit concevoir un nouveau composant intégrant 2 multiplieurs, 1 additionneur, 7 registres et 2 bancs mémoire (voir figure 1 (d)). Le concepteur peut utiliser des composants préconçus ou des macro-générateurs, qui reposent toutefois sur des modèles architecturaux fixes avec des possibilités de personnalisation restreintes.

Un nouveau flot de conception basé sur la synthèse HLS sous contraintes est nécessaire pour introduire de la flexibilité et faciliter la conception de composants orientés TDSI. Il doit inclure (1) des modèles formels pour représenter les contraintes d'E/S et de mémorisation, (2) des phases d'analyse pour vérifier la faisabilité en fonction des contraintes et (3) des méthodes et techniques pour optimiser la synthèse.

## 3. Approche de conception

Le point d'entrée de notre outil de synthèse est une description algorithmique qui spécifie l'algorithme d'une fonction du domaine du TDSI. La spécification est d'abord compilée pour obtenir une représentation intermédiaire, le Graphe Flot de Signaux (*SFG*) (voir Figure 1).

### 3.1 Prise en compte des contraintes temporelles

Nous générons un graphe de contraintes algorithmiques (*ACG*) à partir des temps de traversée des opérateurs et des dépendances de données exprimées dans le *SFG*. Les latences des opérateurs sont assignées aux nœuds opération du graphe *ACG* pendant l'étape de sélection de la synthèse de haut niveau. A partir de la description du système et de son modèle architectural, le concepteur connecte le composant au reste du système en spécifiant les cadences des entrées et sorties, les séquences d'accès et les informations temporelles sur les transferts. Nous définissons un modèle formel appelé *IOCG* (IO Constraint Graph) qui supporte l'expression des contraintes d'intégration pour chaque bus ou port qui au travers desquels le composant communique avec son environnement.

Enfin, nous générons un Graphe de Contrainte Globale (*GCG*) en fusionnant les graphes *ACG* et *IOCG*. La fusion est réalisée intégrant les nœuds de *IOCG* et leurs contraintes sur les nœuds d'entrées et de sorties du *ACG*. Une contrainte minimum sur les nœuds de sortie (date au plus tôt pour le transfert de donnée) est transformée en une contrainte de temps maximum sur le *GCG* (date au plus tard pour le calcul et la production d'une donnée).

Nous vérifions dans une première étape que la cadence d'itération spécifiée pour l'application est compatible avec les dépendances de données de l'algorithme et les latences des opérateurs pour la technologie spécifiée. Nous vérifions ensuite que les dates de production des sorties sont toujours respectées pour le jeu de contraintes temporelles fournies pour les données d'entrée.

### 3.2 Prise en compte de la mémoire interne

Dans notre approche, une table mémoire est extraite à partir du *SFG*. Tous les nœuds de données du *SFG* (hors E/S) sont présents dans la table. Le concepteur peut distribuer et placer les données de l'application et ainsi définir un mapping mémoire. A partir de la table, nous construisons un Graphe de Contraintes Mémoire, *MCG* qui représente tous les conflits d'accès aux données. Le *MCG* est ensuite utilisé pour résoudre les conflits d'accès aux données placées en mémoires internes lors de l'ordonnancement des opérations. Des informations complémentaires sur les modèles formels et la conception des mémoires peuvent être trouvées dans [7], [8] et [9].

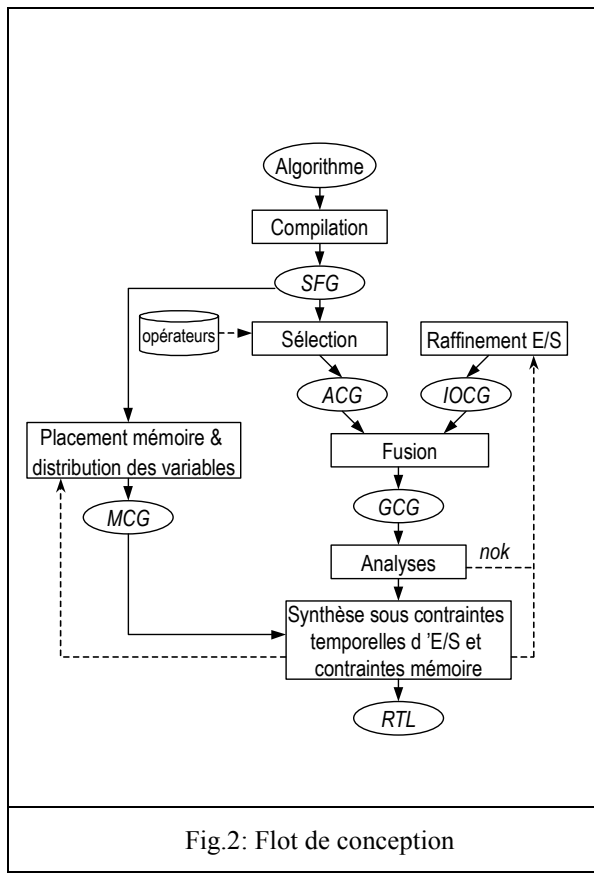

```
Ordonnancement
1)  Calcul de la mobilité(GCG)
2)  pour (temps = 0; temps < fin; temps = temps + t_cycle)
3)    liste = lister les opérations par priorité(GCG)
4)    Ops_exe = trouver les opérations executables (Liste, temps)
5)    assigner(Ops_exe, ens_operateurs, MCG, temps)
6)  fin pour

assignation
1)  tant que (Ops_exe!= NULL)
2)    Ops_faible_mobilité = prendre-premier(Ops_exe)
3)    if(Op_faible_mobilité->marge > 0)
4)      If(Trouver_mem_conflic(MCG, Ops_faible_mobilité) = FALSE)
5)        If(ens_opérateurs != NULL)
6)          Assignation_Ops(sh_liste, opérateur)
7)        else //pas de conflit de resources ni d'accès à la mémoire
8)          Retarder(Ops_faible_mobilité)
9)    else // marge = 0
10)     If(Trouver_conflit_mem (MCG, Ops_faible_mobilité) = FALSE)
11)       creation_Opérateur_()
12)       Ops_Assignation(sh_liste, opérateur)
13)     else
14)       Exit(cycle, opérateur, operation, banc_mémoire, …)
15)   end if
16) End while
```

Fig.2: Flot de conception | Fig.3: Pseudo code de l'algorithme d'ordonnancement

## 3.3 Ordonnancement sous contraintes d'E/S et mémoire

Un ordonnancement classique par liste de priorité repose sur une heuristique de type glouton dans laquelle les opérations exécutables (opérations à ordonnancer) sont classées suivant un ordre de priorité. Dans notre outil, un premier ordonnancement est réalisé à partir du *GCG*. Dans cet ordonnancement, la fonction de priorité dépend de la mobilité des opérations ( date ALAP – date ASAP), (voir figure 3). Pour les opérations ayant la même mobilité, le second critère est la marge (date ALAP – temps_courant). Ensuite les opérations sont ordonnancées et assignées aux opérateurs.

Une opération peut être ordonnancée si la date du cycle courant est supérieure à la date ASAP de l'opération. Toutefois, si deux opérations exécutables ont besoin de la même ressource matérielle, l'opération qui a la mobilité la plus faible (donc la priorité la plus forte) est ordonnancée, la seconde est retardée. Lorsque la marge devient nulle, de nouveaux opérateurs sont alloués aux opérations. Pour réaliser un ordonnancement sous contraintes mémoire, nous introduisons des opérateurs d'accès mémoire et utilisons un critère d'accessibilité aux données sur le *MCG*. A chaque cycle d'accès à la mémoire, il y a autant d'accès mémoire possibles qu'il y a de ports d'accès libres. Les opérations exécutables sont toujours classées suivant la mobilité, mais les opérations dont les opérandes ne respectent pas le critère d'accessibilité à la mémoire sont retirées de la liste. Ainsi, lorsque la marge est égale à zéro et que les opérations ne peuvent être ordonnancées à cause des conflits d'accès à la mémoire, le processus de synthèse est arrêté et le concepteur doit cibler une solution alternative pour l'architecture du composant matériel en révisant le mapping mémoire ou en modifiant les caractéristiques des communications.

Notre technique d'ordonnancement est illustrée en figure 3 et reprend l'exemple précédent. Les contraintes sont maintenant les suivantes : $S = (a|b,c)$. La table mémoire (figure 4 (a) est extraite à partir du *SFG*. Le concepteur a défini un mapping mémoire dans lequel les données sont distribuées et placées dans un seul banc mémoire. Les accès à cette mémoire peuvent être séquentiels (1 cycle) ou aléatoires (2 cycles). Les données internes *var1* et *var2* sont respectivement placées aux adresses @1 et @0 dans le banc mémoire *bank0*. Notre outil construit le *MCG* (figure 4 (b)) correspondant. En plus des contraintes mémoire, le concepteur spécifie deux latences $Lat1=3$ cycles et $Lat2=2 cycles$.

**Pour la latence *Lat1*,** les accès séquentiels s'effectuent entre $var2 \rightarrow var1$ (un arc pointillé avec le poids Wseq entre $var2 \rightarrow var1$). Pour gérer les conflits d'accès à la mémoire, nous définissons un tableau d'accès pour chaque port. Dans notre exemple, le tableau n'a qu'une ligne puisque nous utilisons une seule mémoire simple port (bank0). Le tableau des accès mémoire contient N éléments, avec N = cadence/temps d'accès mémoire séquentielle. La valeur de chaque élément du tableau indique si la mémoire est accessible ou non au temps de cycle courant(c_step). Nous utilisons le *MCG* pour produire un ordonnancement qui favorise les accès en rafale à la mémoire. Si deux opérations ont la même priorité (margin = Lat1-T(+)-T(*) = 1 cycle) et requièrent un accès à la même mémoire, l'opération qui sera ordonnancée sera celle dont l'adresse de l'opérande est consécutive à celle du précédent accès. Par exemple, les opérations de multiplication *(a*var1)* et *(b*var2)* ont la même mobilité. Au cycle c_step *cs_1*, elles sont toutes deux exécutables et leurs opérandes *var1* et *var2* sont en dans le banc mémoire bank0. Le *MCG_1* indique que la séquence *var2* → *var1* plus courte que *var1* → *var2*. Nous ordonnançons *(b\*var2)* au cycle c_step *cs_1* et *(a\*var1)* au cycle c_step *cs_2* pour favoriser les accès (voir figure 4 (c)).

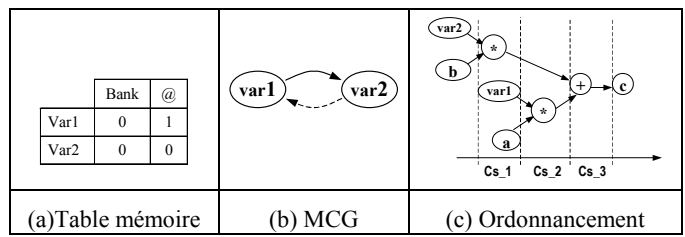

Fig. 4: Scheduling under I/O timing and latency constraint

**Pour la latence *Lat2*,** les opérations de multiplication *(a\*var1)* et *(b\*var2)* ont la même mobilité (mobilité nulle). Les deux opérations doivent obligatoirement être ordonnancées au cycle c_step *cs_1*. A cause des conflits d'accès à la mémoire, il n'y a pas de solution d'ordonnancement. Le concepteur doit donc revoir les contraintes de conception. Il peut cibler une solution alternative en ajoutant un banc mémoire ou en augmentant la latence du traitement.

## 4. Résultats expérimentaux

Nous présentons les résultats de synthèse sous contraintes utilisant notre outil de synthèse de haut niveau GAUT [10]. L'algorithme utilisé pour ces expériences est une FFT. Cette FFT lit 128 échantillons réels *X(k)* en entrée et produit des sorties *Y(k)* composées d'une partie réelle *Yr(k)* et d'une partie imaginaire *Yi(k)*. Le *SFG* de cet algorithme est composé de 16897 arcs et 8451 nœuds. Plusieurs synthèses sont réalisées en utilisant une fréquence d'horloge à 200 MHz et une bibliothèque dont les temps de traversée des opérateurs sont de 2 cycles pour les multiplieurs et de 1 cycle pour les additionneurs, les soustracteurs et les accès mémoire.

Dans la première expérience, nous synthétisons le composant FFT sous contrainte d'E/S, puis analysons les besoins mémoire. La latence de la FFT est définie par la contrainte temporelle maximale entre le premier nœud d'entrée et le dernier nœud de sortie du *GCG*. La latence spécifiée (la plus courte en considérant les dépendances de données et le temps de traversée des opérateurs) est de 261 cycles. Les contraintes d'E/S impose au composant FFT une seule lecture d'échantillon *X* et une seule production d'échantillon *Y* par cycle d'horloge. Le composant résultant est composé de 20 multiplieurs, 8 additionneurs et 10 soustracteurs (Voir Table 1, Exp1). Pour respecter les contraintes d'E/S, l'architecture mémoire est relativement complexe (8 bancs mémoire).

Dans la deuxième expérience, la synthèse est réalisée uniquement sous contraintes mémoire. C'est le parallélisme d'accès aux données qui fixe la latence minimale du composant. Nous réalisons la synthèse en allouant le même nombre d'opérateurs que pour la première expérience et analysons la latence obtenue et les besoins en bus d'E/S. Les contraintes mémoires sont les suivantes, deux mémoires SRAM simple port sont utilisées pour stocker respectivement les 128 coefficients réels et les 128 coefficients imaginaires. La latence imposée par le mapping mémoire est de 215 cycles (Table 1, Exp 2). Dans ce cas, l'architecture nécessite 50 bus (ports) d'E/S ce qui complique fortement les échanges avec le système. L'ajout d'une unité de communication est alors nécessaire.

La dernière expérience synthétise la FFT sous contraintes d'E/S et sous contraintes mémoire. Nous conservons le mapping mémoire de l'expérience 2 et définissons la latence permettant de garantir la cadence des E/S de la première expérience. L'architecture contient 17 multiplieurs, 8 additionneurs et 10 soustracteurs (Table 1, Exp 3). La latence est plus grande mais la complexité de l'architecture est moindre puisqu'il n'y a que deux bancs mémoire et que l'introduction d'une unité de communication n'est pas nécessaire. La synthèse sous contraintes mémoire et contraintes d'E/S permet de gérer efficacement les communications et la mémorisation en conservant une complexité d'architecture raisonnable.

Table 1 : Résultats de synthèse

| | Banc mémoire. | Bus d'entrée | Bus de sortie | Sous. | Add. | Mult. | latence (en cycle) |
|---|---|---|---|---|---|---|---|
| **EXP1** | 8 | 1 | 2 | 10 | 8 | 20 | 261 |
| **EXP2** | 2 | 36 | 14 | 10 | 8 | 20 | 215 |
| **EXP3** | 2 | 1 | 1 | 10 | 8 | 17 | 343 |

## 5. Conclusion

Dans ce papier, une méthodologie de conception de composants matériels sous contraintes d'E/S et mémoire est présentée pour les applications TDSI. Elle repose sur la modélisation des contraintes, leurs analyses et la synthèse de haut niveau. Elle aide le concepteur à implanter efficacement des applications complexes sur des composants matériels. Les résultats produits pour une application TDSI montrent l'intérêt de la méthodologie et de la modélisation des contraintes qui permettent de réaliser un compromis entre la latence, la cadence des entrées sortie et le mapping mémoire.

## References

[1] Codesimulink, http://polimage.polito.it/groups/codesimulink.html
[2] J. Ruiz-Amaya, and Al., " *MATLAB/SIMULINK-Based High-Level Synthesis of Discrete-Time and Continuous-Time Modulators*", In Proc. of DATE 2004.
[3] L. Reyneri, F. Cucinotta, A. Serra, and L. Lavagno. « A hardware/software co-design flow and IP library based on Simulink. », In Proc. of DAC, 2001.
[4] Mathworks, http://www.mathworks.com/
[5] H. Ly, D. Knapp, R. Miller, and D. McMillen, « Scheduling using behavioral templates," in *Proc. Design Automation Conference DAC'95*, June 1995
[6] D. Knapp, and *al.*, « Behavioral synthesis methodology for HDL-based specification and validation," in *Proc. of DAC*, 1995.
[7] P. Coussy, E. Casseau, P. Bomel, A. Baganne, E. Martin*, "Constrained Algorithmic IP Design for System-On-Chip", in Integration, the VLSI journal, 2004 (to appear)*.
[8] P. Coussy , D. Gnaedig, and al., " A Methodology for IP Integration into DSP SoC: A Case Study of a MAP Algorithm for Turbo Decoder"*, In Proc. of ICASSP, 2004*
[9] G. Corre, E. Senn, and al., "Memory accesses Management During High Level Synthesis", *In Proc. of CODES-ISSS, 2004.*
[10] GAUT tool: http://web.univ-ubs.fr/gaut/